\documentstyle[11pt]{article}
\begin{document}
\thispagestyle{empty}
\baselineskip 18pt
\rightline{UOSTP-99-005}
\rightline{SNUTP-99-031}
\rightline{KIAS-P99042}
\rightline{{\tt hep-th/9906119}}
%j3rhf

\

\def\tr{{\rm tr}\,} \newcommand{\beq}{\begin{equation}}
\newcommand{\eeq}{\end{equation}} \newcommand{\beqn}{\begin{eqnarray}}
\newcommand{\eeqn}{\end{eqnarray}} \newcommand{\bde}{{\bf e}}
\newcommand{\balpha}{{\mbox{\boldmath $\alpha$}}}
\newcommand{\bsalpha}{{\mbox{\boldmath $\scriptstyle\alpha$}}}
\newcommand{\bbeta}{{\mbox{\boldmath $\beta$}}}
\newcommand{\bgamma}{{\mbox{\boldmath $\gamma$}}}
\newcommand{\bsbeta}{{\mbox{\boldmath $\scriptstyle\beta$}}}
\newcommand{\blambda}{{\mbox{\boldmath $\lambda$}}}
\newcommand{\bphi}{{\mbox{\boldmath $\phi$}}}
\newcommand{\bslambda}{{\mbox{\boldmath $\scriptstyle\lambda$}}}
\newcommand{\ggg}{{\boldmath \gamma}} \newcommand{\ddd}{{\boldmath
\delta}} \newcommand{\mmm}{{\boldmath \mu}}
\newcommand{\nnn}{{\boldmath \nu}}
\newcommand{\diag}{{\rm diag}}
\newcommand{\bra}[1]{\langle {#1}|}
\newcommand{\ket}[1]{|{#1}\rangle}
\newcommand{\sn}{{\rm sn}}
\newcommand{\cn}{{\rm cn}}
\newcommand{\dn}{{\rm dn}}

\

\vskip 0cm
\centerline{\Large\bf Low Energy Dynamics for 1/4 BPS Dyons}

\vskip 0.2cm

\vskip 1.2cm
\centerline{\large\it
Dongsu Bak $^a$\footnote{Electronic Mail: dsbak@mach.uos.ac.kr},
Choonkyu Lee $^b$\footnote{Electronic Mail: cklee@phya.snu.ac.kr},
Kimyeong Lee $^b$\footnote{Electronic Mail: kimyeong@phya.snu.ac.kr},
and Piljin Yi $^c$\footnote{Electronic Mail: piljin@kias.re.kr}}
\vskip 10mm
\centerline{ \it $^a$ Physics Department,
University of Seoul, Seoul 130-743, Korea}
\vskip 3mm
\centerline{ \it $^b$ Physics Department and Center for Theoretical
Physics}
\centerline{ \it Seoul National University, Seoul 151-742, Korea}
\vskip 3mm
\centerline{ \it $^c$ School of Physics, Korea Institute for
Advanced Study}
\centerline{\it
207-43, Cheongryangri-Dong, Dongdaemun-Gu, Seoul 130-012, Korea}

\vskip 1.2cm
\begin{quote}
{\baselineskip 16pt Classical 1/4 BPS configurations consist of 1/2
BPS dyons which are positioned by competing static forces from
electromagnetic and Higgs sectors. These forces do not follow the
simple inverse square law, but can be encoded in some low energy
effective potential between fundamental monopoles of distinct
types. In this paper, we find this potential, by comparing the exact
1/4 BPS bound from a Yang-Mills field theory with its counterpart
derived from low energy effective dynamics of monopoles. Our method is
generalized to arbitrary gauge groups and to arbitrary BPS
monopole/dyon configurations. The resulting effective action for 1/4
BPS states is written explicitly, and shown to be determined entirely
by the geometry of multi-monopole moduli spaces.  We also explore its
natural supersymmetric extension.}
\end{quote}

%\pacs{14.80.Hv,11.27.+d,14.40.-n}

\newpage

\section{Introduction}

Recently there have been much activity in trying to understand  the
nature of 1/4 BPS dyonic configurations in N=4 supersymmetric Yang-Mills
theories~\cite{yi,hash,bak}. The N=4 supersymmetric theories arise as an
effective low energy theory of parallel D3 branes in the type IIB string
theory~\cite{witten}. The expectation values of the six Higgs fields
are the coordinates of these D3 branes in the transverse six
space. When D3 branes lie on a line, or all Higgs expectation
values are aligned in Lie algebra, there can be only 1/2 BPS
configurations. Most general 1/2 BPS states are collections of 1/2
BPS dyons whose electric charges are all proportional to the
magnetic charge, individually as well as collectively. When the
expectation values of the Higgs fields are not aligned, on the other
hand, there can be 1/4 BPS configurations, which have a nice string
interpretation as multi-pronged strings~\cite{bergman}. As a field
theoretic solution, a 1/4 BPS configuration can be thought of as more
than one 1/2 BPS dyons at rest with respect to each other; the positions
of the component dyons are determined by a delicate balance of the
electromagnetic Coulomb and Higgs forces~\cite{yi}. Because of
this, the relative electric charges of distinct-type dyons are
functions of their relative positions.

The low energy dynamics of 1/2 BPS monopoles has been explored before
in many directions, but only in the context of aligned vacua where  no
static forces among monopoles are possible. The 1/2 BPS
configurations are characterized by their moduli parameters, and
Manton proposed that the low energy motion of 1/2 BPS monopoles be
treated as the geodesic motion on the moduli space\cite{manton}.  There are
several explicitly known moduli space metrics. For a pair of identical
monopoles in SU(2) gauge theory, the moduli space is the
Atiyah-Hitchin manifold\cite{atiyah}. For a pair of distinct
monopoles in SU(3) theory, it is the Taub-NUT manifold~\cite{connell},
and so on.

On the other hand, the low energy dynamics of BPS solitons in
misaligned vacua has been more problematic. In the simplest
example of SU(3) gauge theory, a 1/4 BPS configuration is known to
consist of two 1/2 BPS dyons that are separated by a fixed distance,
at which the Coulomb repulsion due to the relative electric charge is
exactly balanced against a static attraction induced by Higgs
interaction. The explicit form of the potential for the Higgs force,
however, has not been well understood.

Recall that the low energy dynamics of BPS solitons explores physics
that deviates a little bit from the BPS bound. Thus, in certain limit
where 1/4 BPS states are almost 1/2 BPS, it should be possible to
rediscover physics of 1/4 BPS configurations from the dynamics of 1/2
BPS states. Since we have static forces between 1/2 BPS solitons in
misaligned vacua, the simplest possibility is to add a potential term
to the moduli space dynamics. In this paper, we find such a low energy
effective action that describes both the 1/2 BPS and the 1/4 BPS
configurations in misaligned vacua.

The BPS equations for the 1/4 BPS configurations can be grouped into
two sets of equations~\cite{yi}; the first is the old 1/2 BPS
equations that produce purely magnetic monopoles, and the second
solves for the unbroken global gauge modes in this magnetic
background. The solution of the second BPS equation is guaranteed and
determines electric charges carried by monopoles of the first BPS
equation. Thus, the 1/4 BPS dyons are constructed by dressing 1/2 BPS
monopoles electrically, where the amount of the relative electric
charge depends on the monopole moduli parameters. A crucial
consequence is that the moduli space of monopoles also parameterizes
the classical 1/4 BPS dyons but with a twist that some of the
parameters characterizes  electric charges.

These observations tell us that there are two different ways of
constructing 1/4 BPS configurations. The first is to obtain an exact
field theoretic classical solution. The energy of such a configuration
would saturate the classical BPS bound exactly. On the other hand,
since 1/4 BPS dyons are all dressed versions of purely magnetic
monopoles, one also should be able to find them as excited charged
configurations on the moduli space dynamics.  They should also
saturate a BPS bound of the low energy effective dynamics. In the
limit where the moduli space approximation is good, one then
identifies these two BPS bounds, which should constrain the unknown
potential term.  As a matter of fact, this procedure turns out to be
enough to fix  the potential completely.

We will first consider the interaction of two distinct dyons in the
SU(3) case. We find the exact form of the potential, using the idea
outlined above.  Furthermore, the resulting Hamiltonian of the low
energy dynamics is shown to have a BPS bound.  As a consistency check,
we show that the configurations that saturate this low energy BPS
bound describe the identical physics as the field theoretic 1/4 BPS
configurations. We also show that the $1/r$ piece of the potential at
large separation is consistent with the results from the study of the
interaction of two point-like dyons in large separation.

We then  generalize this discussion  to any  combination of magnetic
monopoles in arbitrary gauge group. The form of the potential will
turn out to be half the norm of of certain triholomorphic Killing
vector field on the moduli
space. Here a recent work by D. Tong \cite{tong} plays a crucial role.
This way, the effective Lagrangian is again determined by the geometry
of the moduli space alone. This particular form of potential is known
to admit supersymmetric extension, which we also explore.

The plan of the paper is as follows. In Sec.2, we review briefly the
1/4 BPS configuration of two distinct dyons in SU(3) gauge theory and
the moduli space metric of these dyons in the 1/2 BPS limit. In Sec.3,
we obtain the exact potential for the simplest case of $SU(3)$ when
the deviation from the 1/2 BPS case is small, that is, when the D3
branes are almost on a straight line. We show that the BPS
configurations of the low energy dynamics are identical to the BPS
configuration of the field theoretic ones.  In Sec.4, we generalize
this result to arbitrary monopole configurations in arbitrary gauge
group with more emphasis on the geometrical character of the low
energy effective action.  The underlying supersymmetric Lagrangian and
supercharge, as well as the BPS conditions, are found in Sec. 5.  In
Sec. 6 we conclude with some remarks.

\section{Two Distinct Monopoles in the SU(3) Gauge Theory}

The SU(3) gauge group appears in the low energy dynamics of three
parallel D3 branes. (See Ref.~\cite{yi} for details.)  The positions
of D3 branes on a plane of the six-dimensional transverse direction
are dictated by the expectation values of two Higgs fields:
\beqn
&& \hat b\cdot\phi(\infty)=  \diag (h_1,h_2,h_3), \label{basy} \\
&& \hat a\cdot\phi(\infty)=  \xi\; \diag (h_1,h_2,h_3)+ \eta\;
\diag (\mu_2,-\mu_1-\mu_2,\mu_1) , \label{aasy}
\eeqn
where $h_1<h_2<h_3$, $h_1+h_2+h_3=0$ and $\mu_1=h_2-h_1,
\mu_2=h_3-h_2$. (Here the string tension is multiplied to the positions
of D3 branes so that they acquire the mass dimension.) Two Higgs field
expectation values are the coordinates of D3 branes along two
orthogonal directions ${\hat b} = (1,0)$ and ${\hat a}= (0,1)$. The
relative position vector of the second D3 brane with respect to the
first D3 brane on the plane is
\beq
{\vec R}_1 = (\mu_1, \xi \mu_1 - \eta(\mu_1+2\mu_2)),
\label{R1}
\eeq
while  the relative position of the third D3 brane with respect to the
second D3 brane is
\beq
{\vec R}_2 = (\mu_2, \xi\mu_2+\eta(2\mu_1+\mu_2)).
\label{R2}
\eeq

Two simple roots  $\balpha$ and $\bbeta$ of the  $SU(3)$ group are
chosen in accordance with the convention $\balpha^2=\bbeta^2=1$,
$\balpha\cdot \bbeta = -1/2$.   With $\balpha\cdot {\bf H} =
\frac{1}{2} \diag (-1,1,0)$ and $\bbeta\cdot{\bf H} =
\frac{1}{2}\diag(0,-1,1)$, the masses of isolated $\balpha$ and
$\bbeta$ monopoles are
\beqn
&& m_1 = g |{\vec R}_1|  ,\\
&& m_2 = g|{\vec R}_2| ,
\eeqn
respectively, where $g = 4\pi$ is the charge of magnetic monopole with
$e=1$ assumed for convenience.  Notice that when $\xi$ and $\eta$ are very
small,
\beqn
&& m_1 = g\mu_1 + \frac{g}{2\mu_1}(\xi \mu_1 -\eta(\mu_1+2\mu_2))^2, \\
&& m_2 = g\mu_2 + \frac{g}{2\mu_2}(\xi \mu_2 + \eta(2\mu_1+\mu_2))^2.
\eeqn

There is a third monopole corresponding to the third positive root
$\balpha + \bbeta$ and its mass is $m_3 = g|\vec{R}_1+
\vec{R}_2|$. In contrast to the case when the Higgs vacuum values are
aligned, generically there is no distinction between fundamental or
composite monopoles. (For example, consider the case where three D3
branes lie on the corners of an equitriangle.)  However, when the
Higgs vacuum values are almost aligned as in the case we study, we can
still distinguish between fundamental and composite monopoles. Thus,
$\balpha$ and $\bgamma$ monopoles are fundamental and
$\balpha+\bgamma$ monopoles are composite.

The dyonic configuration we consider is made of one $\balpha$
and one $\bbeta$ monopoles, with electric charges $q_1$ and $q_2$,
respectively. Thus the asymptotic forms of the Higgs fields are
\beqn
&& \hat b\cdot\phi = \hat b\cdot\phi(\infty)  -  \frac{1}{4\pi  r} \, g
(\balpha +\bbeta)\cdot {\bf H} ,
\label{asymp1} \\
&& \hat a\cdot\phi = \hat a\cdot\phi(\infty) - \frac{1}{4\pi r}\, (q_1
\balpha +q_2\bbeta) \cdot {\bf H} .\label{asymp2}
\eeqn
For the given asymptotic (\ref{asymp1}), the solution of the
first BPS equation
\beq
B_i = D_i \hat b\cdot \phi
\label{bpse1}
\eeq
is uniquely characterized by the relative distance, $L$, between two
monopoles. Once the first BPS solution is found, the solution of
the second BPS equation
\beq
D_i^2 \hat a\cdot\phi - [\hat b\cdot\phi, [\hat b\cdot\phi,\hat
a\cdot\phi]] =0
\label{bpse2}
\eeq
is found to be unique  for a given asymptotic (\ref{asymp2}).  From
this solution, we can read  electric
charges
\beqn
&& q_1 = g  (\xi +\eta p_1) ,\\
&& q_2 = g  (\xi+\eta p_2) ,
\eeqn
where
\beqn
&& p_1 = \frac{\mu_1 -\mu_2-2(\mu_1+2\mu_2)\mu_2 L}{\mu_1+\mu_2+
2\mu_1\mu_2 L}, \\
&& p_2 = \frac{\mu_1-\mu_2+2(2\mu_1+\mu_2)\mu_1
L}{\mu_1+\mu_2+2\mu_1\mu_2 L}.
\eeqn

As discussed in the introduction, one  of the interesting things about
these BPS solutions  is that we
may treat  the 1/4 BPS configurations as if they are made of  1/2 BPS
monopoles with some electric
dress. The  first BPS equation can be regarded as the 1/2 BPS
equation. Its solution describes a collection of
 1/2 BPS monopoles and is  characterized uniquely by the moduli space
coordinates.
Here the constituent monopoles carry mass characterized by the Higgs
asymptotic value $\hat b\cdot\phi(\infty)$. Then we solve the second
BPS  equation, which is identical to the gauge zero mode equation of the
first BPS equation. The solutions of the second BPS equation is
determined uniquely by the moduli parameters of the  solution of the
first BPS equation and by the value $\hat{a}\cdot \phi(\infty)$. One of the
key point  of this paper is to take this view further and to regard
the low energy dynamics of 1/4 BPS dyons as that of 1/2 BPS monopoles
described by the first BPS equation.

Thus, instead of $m_1$ and $m_2$, we regard $g\mu_1$ and $g\mu_2$ as
the real mass of monopoles. We
can define the total and relative charges with respect to the
masses $g\mu_1$ and $g\mu_2$, that is,
\beqn
&& q_{\rm tot} = \frac{\mu_1q_1+\mu_2q_2}{\mu_1+\mu_2}= g \left( \xi +
\eta \frac{\mu_1-\mu_2}{\mu_1+\mu_2} \right) \label{qtot}, \\
&& \Delta q(L) = \frac{q_2-q_1}{2} = \frac{\Delta q_c}{1+
\frac{1}{2\mu L}},
\label{dqr}
\eeqn
where   the critical relative charge is defined as
\beq
\Delta q_c \equiv g\eta \frac{\mu_1^2+\mu_1\mu_2 +\mu_2^2}{\mu_1\mu_2},
\label{dqc}
\eeq
and the pseudo relative mass is
\beq
\mu = \frac{\mu_1\mu_2}{\mu_1+\mu_2}.
\eeq
Notice that the relative charge vanishes when $L=0$ and
tends  to $\Delta q_c$ at $L=\infty$. (When three D3 branes lie on a line
and so $\eta=0$, the relative charge vanishes.) We further
note that the mass difference, $\sum_i(m_i- g\mu_i)$, may be
 written, in terms of charges, as
\beq
m_1+m_2-g(\mu_1+\mu_2)= {\mu_1+\mu_2\over 2g}(q_{\rm tot})^2 +
{2\mu\over  g} (\Delta q_c)^2.
\label{difference}
\eeq
 The difference between
$m_i$ and $g\mu_i$ can be considered as a result of the interaction
between monopoles, which even includes a constant potential.

\subsection{BPS Energy}

In the BPS energy bound, it is natural to introduce
the two-dimensional magnetic and electric  charge vectors,
\beqn
&& {\vec Q}^M = g( {\vec R}_1 + {\vec R}_2), \\
&& {\vec Q}^E = q_1 {\vec R}_1 + q_2 {\vec R}_2 .
\eeqn
The central charge of the  N=4 supersymmetric algebra, which gives the
BPS energy bound, is
\beq
Z^2_\pm  = {\rm max} \left( {\vec Q}_M^2 + {\vec Q}_E^2 \pm 2{\vec Q}_E\times
{\vec Q}_M \right) .
\eeq
When the $BPS$ equations are  satisfied so that the BPS energy bound
is saturated,  there is another expression for
the central term $Z$. Since ${\hat a}\cdot {\vec Q}^M = {\hat b}\cdot
{\vec Q}^E$, the energy is
\beqn
Z_+ &=&  {\hat b}\cdot {\vec Q}^M + {\hat a}\cdot {\vec Q}^E
\nonumber    \\
 &=&  g(\mu_1+\mu_2) +\frac{\mu_1+\mu_2}{g} (q_{\rm tot})^2
 + \frac{4\mu }{g}  \Delta q(L) \Delta q_c,
\label{bps0}
\eeqn
which is exact. As  ${\hat b}\cdot {\vec Q}^M= g(\mu_1+\mu_2) $ is
just the sum of constituent monopole masses, we regard  the rest of
the contribution, ${\hat a}\cdot {\vec Q}^E$, to arise from the
dynamics of monopoles.

The low energy dynamics means that the dynamical energy contribution
to the rest mass is small. From Eqs.~(\ref{qtot}) and (\ref{dqr}), we
see that the low energy approximation may hold if
\beq |\xi|,\;\; |\eta|\; << 1 . \eeq
{}From Eqs.~(\ref{basy}) and (\ref{aasy}), we see that three D3 branes
are almost collinear, for the low energy approximation to hold. The
above condition also implies that the magnitude of $q_{\rm tot},
\Delta q$ of 1/4 BPS configurations should be very small compared with
the magnetic charge $g$.  {}From our point of view, ${\hat a}\cdot
{\vec Q}^E$ in the central charge arises from the low energy dynamics
of monopoles. As we will see, it has the contributions from both
kinetic and potential energies.

\subsection{Moduli Space Metric of 1/2 BPS Monopoles}

When three D3 branes are collinear, the low energy dynamics of two
monopoles can be described by the moduli space or collective
coordinate dynamics.  There are four zero modes for each monopole,
three for its position and one for the U(1) phase. We call their
positions and phases to be ${\bf x}_i,\psi_i, i=1,2$ for $\balpha$ and
$\bbeta$ monopoles, respectively.  The exact nonrelativistic effective
Lagrangian has been found to be a sum of the Lagrangians for the
center of mass and the relative motion~\cite{connell}.  As there is no
external force, the center of mass Lagrangian is free one,
\beq
{\cal L}_{\rm cm} = \frac{g(\mu_1+\mu_2)}{2} \dot{{\bf X}}^2 +
\frac{g}{2(\mu_1+\mu_2)} \dot{\chi}^2 ,
\label{cmL}
\eeq
where the center of mass position is ${\bf X} = ( \mu_1 {\bf x}_1 +
\mu_2 {\bf x}_2)/(\mu_1+\mu_2)$ and the center of mass phase is $\chi
=\psi_1+\psi_2 $.
The relative motion between them is more complicated and described
by the Taub-NUT metric,
\beq
{\cal L}_{\rm rel}
= \frac{g\mu }{2} \left( ( 1+ \frac{1}{2\mu r}) \dot{\bf r}^2
+ \frac{1}{4\mu^2(1+ \frac{1}{2\mu r})}
(\dot{\psi} + {\bf w}({\bf r})\cdot \dot{\bf r})^2 \right),
\label{relL}
\eeq
where the relative position is ${\bf r} = {\bf x}_2-{\bf x}_1$, the
relative phase is $\psi =2(\mu_1\psi_2-\mu_2\psi_1)/(\mu_1+\mu_2)$,
and ${\bf w}({\bf r}) $ is the Dirac potential such that $\nabla
\times {\bf w}({\bf r}) = -{\bf r}/r^3$. The range of $\psi$ is
$[0,4\pi]$. The relative moduli space ${\cal M}_0$ is the Taub-NUT
manifold with the metric given from the above Lagrangian.
The eight-dimensional moduli space is then given by
\beq
{\cal M} = R^3 \times \frac{ S^1 \times {\cal M}_0}{Z},
\eeq
where $Z$ is the identification map
\beq
(\chi, \psi) = (\chi + 2\pi, \psi + \frac{4\pi \mu_2}{\mu_1+\mu_2}).
\eeq

One way of obtaining this moduli space metric is by exploring the long
range interaction between two dyons\cite{manton1}.  As for two dyons
of charge $(g, \pi_1)\balpha$ and $(g,\pi_2)\bbeta$, there are several
interactions between them. Obvious ones are electric and magnetic
Coulomb potentials. Besides, there exists a potential due to Higgs
force between them.. In addition, when they move, there is a time delay
effect which appears in the form of Lienard-Wiechert potential. When
they are put together, we get the Routhian obtained from the Legendre
transformation of phase variables to charges,
\beq
{\cal R}=  \frac{g(\mu_1+\mu_2)}{2} \biggl(\dot{\bf X}^2 -
\frac{(\pi_{\rm t})^2}{g^2} \biggr)
+ \frac{g\mu}{2} \bigl( 1+ \frac{1}{2\mu r}\bigr) \biggl(\dot{\bf r}^2 -
\frac{4(\pi_{\rm r})^2}{g^2} \biggr) +\pi_{\rm r}
{\bf w}({\bf r})\cdot \dot{\bf r},
\label{routhian}
\eeq
where the relation between $\pi_{\rm t}, \pi_{\rm r}$ and
$\pi_1$ and $\pi_2$ are given by similar equations as in
Eqs.~(\ref{qtot}) and (\ref{dqr}).  We emphasize
here that total charge $\pi_{\rm t}$ and relative charge $\pi_{\rm r}$
are conjugate momenta without any fixed value.

\section{Case of $SU(3)$}

To find the potentials between two dyons in every relative separation,
we explore first the case $q_{tot}=\pi_{\rm t}=0$. Suppose that the
exact potential for the relative motion is ${\cal U}_{\rm rel}(r)$, and
so the effective potential for two dyons is
\beq
U_{\rm eff}(r) = \frac{2\mu}{g}\left( 1+ \frac{1}{2\mu r}\right)
(\pi_{\rm r})^2 + {\cal U}_{\rm rel}(r)
\label{ueff0}
\eeq
The first part of the effective potential comes from the charge
kinetic energy. This effective potential should have a minimum at $r$
if $\pi_{\rm r} = \Delta q(r)$ for $\Delta q(r)$ specified by
Eq.~(\ref{dqr}) and the energy at the minimum should be identical to
the BPS energy (\ref{bps0}) once we add the mass of monopoles. Then we
realize that the potential energy is exactly one half of the excessive
BPS energy associated with the relative charge. Using this, we find
the piece of the potential,
\beqn
{\cal U}_{\rm rel}(r) &=& \frac{2\mu}{g}
\frac{(\Delta q_c)^2}{ 1 + \frac{1}{2\mu r}} \nonumber \\
&=& \frac{2\mu}{g}( 1 + \frac{1}{2\mu r})(\Delta q(r))^2
\label{pote0}
\eeqn
where $\Delta q (r)$ is to be understood as function of $r$ given in
Eq.~(\ref{dqr}).  However, we have not quite found the potential $\cal
U$. We actually found only the piece that depends on the distance
$r$. By including the center of mass motion, it is not difficult to
guess that the actual potential must be of the form,
\beq
{\cal U}(r)=\frac{\mu_1+\mu_2 }{2g}q_{\rm tot}^2 + \frac{2\mu}{g}
\frac{(\Delta q_c)^2}{ 1 + \frac{1}{2\mu r}}
\label{pote}
\eeq
This guess will be justified below momentarily.

As we have seen, this identification can be made for all possible classical
dyons, which translates to all possible values of $r$. In effect,
we have found the potential $\cal U$ throughout the moduli space.
In the next section, we will discover that this method is trivially
generalized to arbitrary configurations of monopoles and dyons in
arbitrary gauge group, and that it produces a rather special kind
of potential, which, as one may expect, admits supersymmetric extension.

Let us go back to the matter concerning total charge and its
associated constant potential energy. If we minimize the effective
potential (\ref{ueff0}), we get $\pi_{\rm r}$ to be $\Delta q(r) =
\Delta q_c/(1+ \frac{1}{2\mu r})$. However, we could not fix $\pi_{\rm
t} $ to be $q_{\rm tot}$ in this way. Hamilton's equations of motion
imply that $\pi_{\rm t}$ to be constant in time. Thus, the naive
equilibrium condition does not fix the total electric charge, even
though it fixes the relative charge.  To understand this, let us now
collect the full low energy Lagrangian for 1/4 BPS dyons, which is the
sum of the kinetic energies (\ref{cmL}), (\ref{relL}) on the moduli
space and the potential energy (\ref{pote})
\beq
{\cal L}_{\rm low} = {\cal L}_{\rm cm} + {\cal L}_{\rm rel} - {\cal U}(r).
\eeq

In terms of the conjugate momenta, ${\bf P}_{\rm t} = g(\mu_1 + \mu_2)
\dot{\bf X}$, ${\bf p} = g\mu (1+ \frac{1}{2\mu r})\dot{\bf r}
+\pi_{\rm r} {\bf w}({\bf r})$, $\pi_{\rm t}$, and $\pi_{\rm
r}$, the Hamiltonian is
\beqn
{\cal H}&= & \frac{1}{2g(\mu_1+\mu_2)} {\bf P}^2 +
\frac{1}{2g\mu}\frac{1}{1+\frac{1}{2\mu r}} ({\bf p}- \pi_{\rm
r}{\bf w}({\bf r}))^2  \nonumber \\
 &  & +\frac{\mu_1+\mu_2}{2g} (\pi_{\rm t})^2+
\frac{2\mu}{g}(1+ \frac{1}{2\mu r}) \pi_{\rm r}^2 + {\cal U}(r) .
\eeqn
The Hamiltonian can be expressed as follows:
\beqn
{\cal H}  &= & \frac{1}{2g(\mu_1+\mu_2)} {\bf P}^2 +
\frac{1}{2g\mu}\frac{1}{1+\frac{1}{2\mu r}} ({\bf p}- \pi_{\rm
r}{\bf w}({\bf r}))^2  \nonumber \\
&  & + \frac{\mu_1+\mu_2}{2g} (\pi_{\rm t} \mp q_{\rm tot})^2+
\frac{2\mu}{g}(1+ \frac{1}{2\mu r}) (\pi_{\rm r}\mp \Delta q(r))^2 \pm
{\cal Z} ,
\eeqn
where the new central term is
\beq
{\cal Z} = \frac{\mu_1+\mu_2}{g} q_{\rm tot} \pi_{\rm t}
 + \frac{4\mu}{g} \Delta q_c \pi_{\rm r}.
\eeq
The central term is linear in the conjugate momenta $\pi_{\rm t}$
and $\pi_{\rm r}$.
Clearly there is a classical BPS bound on the mechanical  energy
\beq
{\cal H} \ge |{\cal Z}|,
\label{cbps}
\eeq
which is saturated for any ${\bf X}$ and ${\bf r}$ when $\pi_{\rm t} =
q_{\rm tot}$ and $\pi_{\rm r} =
\Delta q(r)$ and $\dot{\bf X}=\dot{\bf r} = 0$.  Thus the nonrelativistic
BPS configuration matches exactly the field theoretic BPS
configuration.  In this case, the sum of the rest mass plus this BPS
energy is exactly the 1/4 BPS energy bound (\ref{bps0}).  Thus our 1/4
BPS field configuration corresponds to not the lowest energy
configuration, but a BPS saturated configuration of the nonrelativistic
Hamiltonian.

As an independent check of the above potential, let us consider the
leading term in the large distance. When three D3 branes are not collinear,
two Higgs fields are involved nontrivially and so we expect an additional
Higgs interaction~\cite{fraser}.  To see this, let us go back to the
old derivation of the Higgs interaction. We put the resting  $\balpha$
dyon at ${\bf x}_1$. The Higgs field far from this dyon is
\beq
\phi_I({\bf x})  = \phi_I (\infty)  - \frac{\hat{R}_{1I}}{4\pi
|{\bf x}-{\bf x}_1|} \balpha \cdot {\bf H} \sqrt{g^2 + \pi_1^2},
\eeq
in the unitary gauge. Note that $\balpha \cdot \phi_I(\infty) = \vec{
R}_{1I}$. The $\bbeta$ monopole at ${\bf x}={\bf x}_2$  would feel
this Higgs
field as an effective mass
\beq
 -m_{\rm eff}= - \sqrt{g^2+\pi_2^2}\: \biggl| {\vec R}_2 - {\hat R}_1
\sqrt{g^2+ \pi_1^2} \;\frac{\balpha \cdot \bbeta }{4\pi r} \biggr|,
\label{meff0}
\eeq
where $r=|{\bf x}_2-{\bf x}_1|.$

We expand the effective mass to order $1/r$ and to quadratic terms in $\pi_1$
and $\pi_2$ to get
\beq
-m_{\rm eff} = -\biggl(1+ \frac{\pi_2^2}{2g^2}\biggr) m_2
 - \frac{g^2}{8\pi r} {\hat R}_1 \cdot {\hat R}_2 \biggl( 1+
\frac{\pi_1^2}{2g^2}\biggr)\biggl(  1+ \frac{\pi_2^2}{2g^2} \biggr).
\label{meff}
\eeq
If three D3 branes are  collinear, ${\hat R}_1 = {\hat R}_2$ and
we get the old result.  When we include the velocity dependent terms
and electromagnetic forces  and keep everything in quadratic order, we
get the previous Routhian (\ref{routhian}).

However, when they are not collinear, there is an additional $1/r$
correction to the old result. As we assume that the deviation from the
straight line is very small, or $\eta << 1 $,
\beq
{\hat R}_1 \cdot {\hat R}_2 = \cos \theta \approx 1 -
\frac{\theta^2}{2}.
\eeq
{}From Eqs. (\ref{R1}) and (\ref{R2}), one can show easily that
\beqn
\theta &=& 2 \eta \frac{\mu_1^2+\mu_1\mu_2+\mu_2^2}{\mu_1\mu_2} +{\cal
O}(\eta^2) \nonumber \\
&=&  \frac{2\Delta q_c}{g}
\eeqn
with $\Delta q_c$ in Eq.~(\ref{dqc}).
The small deviation obtained from Eq.~(\ref{meff}) is then
\beqn
\Delta L  &=& -m_1-m_2 +g(\mu_1+\mu_2)+
\frac{g^2}{8\pi r} (1- \hat{ R}_1 \cdot \hat{R}_2) \nonumber
\\ &=&  -{\mu_1+\mu_2\over 2g}(q_{\rm tot})^2 -
{2\mu\over  g} (\Delta q_c)^2 +
 \frac{ (\Delta q_c)^2}{ 4\pi r},
\eeqn
where we have included the additional constant terms from masses that
are given explicitly in (\ref{difference}).  This is the additional
attractive potential which exists since three D3 branes are not
collinear. It is a matter of simple algebra to check that the exact
potential (\ref{pote}) we found above, do contain this term to the
leading $r$-dependent piece at large $r$. One may wonder about terms
of order $1/r^2$. These terms do not come out to be symmetric under
the exchange of $\balpha$ and $\bbeta$ monopoles and depend on the
monopole core size. It needs a further investigation, which we do not
attempt here.

In the above derivation we have dropped terms of order $(\Delta q_c)^2
(\pi_i)^2$ and $(\Delta q_c)^4$. Terms of order $(\Delta q_c)^2
(\pi_i)^2$ can be regarded as the correction to the kinetic term which
exists since D3 branes are not collinear. Corrections of such order is
expected from the naive moduli space metric as there will be
additional term such as $\int d^3x \tr (\hat{a}\cdot \dot{\phi})^2$.
Terms of order $(\Delta q_c)^4$ can be regarded as the correction to
the potential energy and is negligible compare with terms of order
$(\Delta q_c)^2$. 

The low energy approximation we found holds when the kinetic and
potential energies are much smaller than the rest mass. In terms of
generic velocity $v$ and the coefficient, $\epsilon$, of
$\hat{a}\dot{\phi}(\infty)$, the low enegy approximation holds when
\beq
 v<< 1, \;\;\, \epsilon << 1.
\eeq
The order of the low energy Lagrangian is then
\beq
L \sim v^2 + \epsilon^2
\eeq
The corrections we have neglected in the above expansion is then of order
\beq
v^4, \;\;\; v^2 \epsilon^2, \;\;\; \epsilon^4.
\eeq

\section{Generalization}

The crucial ingredient of the preceding section came from a rather
simple observation: we have two different ways of determining  the 1/4
BPS dyon mass. The field theoretic one gives an exact expression,
while the other is derived from approximate low energy dynamics on the
moduli space of monopoles.  This is possible because the moduli space
dynamics already incorporates various internal excitations that induce
relative electric charges. By comparing the two, one obtains an
approximate form of the inter-monopole potential that is arbitrarily
accurate as the Higgs expectations become collinear. In this section,
we will utilize this simple idea to fix the bosonic part of low energy
effective Lagrangian for all 1/2 BPS and 1/4 BPS states in maximally
supersymmetric Yang-Mills theory.

First consider a 1/2 BPS multi-monopole configuration, when an
arbitrary gauge group $G$ is broken to $U(1)^r$ by a single Higgs
field expectation value $\langle {\hat b}\cdot \phi\rangle={\bf h}$ in
the root space.  There are always a set of simple roots $\bbeta_A,
A=1,...r$ such that ${\bf h}\cdot \bbeta_A>0$. Any BPS magnetic
monopole configuration would have the magnetic charge
\beq
{\bf g} = g(n_1 \bbeta_1 + n_2 \bbeta_2+... + n_r \bbeta_r),
\eeq
with nonnegative $n_A's$. Without loss of generality, we
assume that it is irreducible so that
all $n_A$ are positive. In this case all monopoles are interacting
with each other directly or indirectly and are not divided into
noninteracting subclusters.

Each monopole of $\bbeta_A$ type has mass $g\mu_A=g {\bf h}\cdot
\bbeta_A$ and four zero modes. The total energy is then $g\sum_A n_A
\mu_A = {\bf g}\cdot {\bf h} $ and the total number of zero mode is
$4N=4\sum_A n_A$~\cite{ejw}. The low energy dynamics of this
configuration can be described by the dynamics in the moduli space of
dimension $4N$ with the metric $ds^2 = g_{\mu\nu} dz^\mu dz^\nu$. This
moduli space is known to be hyperk\"ahler and thus are equipped with
three covariantly constant complex structures~\cite{atiyah, blum}. The
low energy Lagrangian is then
\beq
{\cal L}_1 = \frac{1}{2}\; g_{\mu\nu}\dot{z}^\mu \dot{z}^\nu .
\label{calL}
\eeq
The unbroken $U(1)^r$ symmetries act on the moduli space as
translational isometries, and generate cyclic $U(1)$ coordinates
$\psi^A$, whose conjugate momenta $q_A$ are conserved electric
charges.  We divide the moduli space coordinates $z^\mu$ into $r$
$\psi^A$'s and remaining $(4N-r)$ $y^i$'s. Up to gauge
transformations, the solution of the first BPS equation (\ref{bpse1})
is uniquely characterized by the values of the coordinates $y^i$.  The
Lagrangian (\ref{calL}) can be rewritten as
\beq
{\cal L}_1 = \frac{1}{2} \; h_{ij} \dot{y}^i \dot{y}^j + \frac{1}{2}
 L_{AB}  (\dot{\psi}^A + w^A_{\; i} \dot{y}^i) (\dot{\psi}^B+ w^B_{\;
j}\dot{y}^j).
\eeq
Due to the cyclic properties of $\psi^A$'s, the metric components
$h_{ij}, L_{AB}$ and $w^A_{\; i}$ are functions of $y^i$ only.
Since the kinetic energy should be positive, the metric $h_{ij}$ and
$L_{AB}$ are positive definite. Each $U(1)$ generators are associated
with the vector field $K_A=\partial/\partial{\psi^A} = \delta_A^\mu
\partial/\partial z^\mu$,
which is a (triholomorphic) Killing vector field. Finally,
note that $L_{AB}(y) = g_{\mu\nu}K_A^\mu K_B^\nu$.

Denoting the conjugate momenta as
\beqn
&& q_A = L_{AB}(\dot{\psi}^B+ w^B_{\; j} \dot{y}^j), \\
&& p_i = h_{ij} \dot{y}^j + q_A w^A_{\; i},
\eeqn
the electric charge of the whole configuration is expressed in terms of
the $q_A$'s as
\beq
{\bf q} = q_1 \bbeta_1 +q_2 \bbeta_2+...+q_r \bbeta_r .
\eeq
The Hamiltonian ${\cal H}_1 = \frac{1}{2} g^{\mu\nu}p_\mu p_\nu$ is
\beq
{\cal H}_1 = \frac{1}{2} h^{ij}(p_i - q_A w^A_{\; i}) (p_j - q_B w^B_{\; j})
   + \frac{1}{2} (L^{-1})^{AB} q_A q_B.
\eeq
Since $L^{-1}$ depends on the $y^i$'s, this shows that  electric
charge excitations will typically induce a long range potential.
Because of this, 1/4 BPS configurations of dyons with relative charge
cannot be allowed if  no other static forces are present,
as would be the case if Higgs expectations were all aligned.

Suppose that the other Higgs field $\hat a\cdot \phi$ acquires an
expectation value which is misaligned with respect to $\langle \hat
b\cdot \phi\rangle$.  Then there is an attractive static force between
monopoles. If the misalignment is small enough so that the static
attractive potential is much smaller than the monopole mass scale, we
may incorporate this potential into the above moduli space
dynamics. (The part of Higgs expectation $\hat a\cdot\phi$ that is
proportional to that of $\hat b\cdot \phi$ is not associated with any
static force. Rather, its effect on the BPS configuration is to fix
the total charge to a certain value and to give an additional
contribution to the energy besides the energy carried by the
charge. This can be understood as the correction to the bare mass of
magnetic monopoles.)

Adding a potential to the Hamiltonian and considering the static
configuration with $\dot{y}_i=0$, we get an effective potential
\beq
{ U}_{\rm eff} = \frac{1}{2} q_A (L^{-1}(y))^{AB} q_B + {\cal U}(y).
\eeq
For 1/4 BPS configurations, $q_A$ should be determined by the
 moduli parameters $y^i$, or reversely, given values of  $q_A$
restrict the equilibrium positions $y^i$.

On the other hand, an exact expression of the 1/4 BPS configuration
is known. The additional energy due to the electric charge is given
by a simple expression,
\beq
E_Q = {\hat a}\cdot {\vec Q}^E = a^A \bar{q}_A(y).
\eeq
The second equality defines the projected Higgs expectation values $a^A$.
The electric charge
are fixed by the positions of magnetic monopoles, which we emphasize
by introducing new notation $\bar{q}_A(y)$. A recent work by
D. Tong~\cite{tong} brought some light on this quantity. He expressed
$\bar{q}_A(y)$ in
terms of quantities on moduli space and found
\beq
\bar{q}_A (y) \equiv L_{AB}(y)\, a^B.
\eeq
This equation can be viewed in two equivalent ways.
One is as a restriction on the
possible equilibrium positions $y^i$ when the electric charges are  given.
Or equivalently, as an expression for  electric charges in terms of
the equilibrium positions $y^i$. Either way, the excess energy due to the
electric charge excitation is
\beq
a^A  \bar{q}_A (y) = a^A L_{AB}(y) \,a^B = \bar{q}_A(y)\,
 (L^{-1}(y))^{AB} \bar{q}_B(y) .
\eeq
Note that this happens to be twice the charge kinetic energy if we put
$q_A= \bar{q}_A(y)$.  Again we demand the potential ${\cal U}(y)$ to
be identical to one  half of $E_Q$, that is, when expressed as a
function of $y^i$,
\beq
{\cal U}(y) = \frac{1}{2}\, a^A L_{AB}(y) \,a^B .
\eeq
We need to pause here for a moment, and explain how it was possible
that we obtained the potential ${\cal U}$ by considering only  the
classical minimum energy configurations of ${U}_{\rm eff}$. It may
seem that we made an extrapolation of some kind.  However, it  is not
the case. We actually have derived the exact potential, as we will
explain below. 

{}From Ref.~\cite{yi}, we know that solutions to the
first BPS equation (\ref{bpse1}) are characterized by moduli
parameters $y^i$'s. That is, any given set of $y^i$, a purely magnetic
solution exists. Then one solves the second BPS equation
(\ref{bpse2}), which leads to 1/4 BPS dyons whose configuration
satisfies the relation $q_A= \bar{q}_A(y) = L_{AB}(y) a^B$. In other
words, there are 1/4 BPS configurations for any generic values of
$y^i$'s.  Therefore, while we identified the value of potential $\cal
U$ for individual 1/4 BPS states, we can learn the values of $\cal
U$ for all $y^i$ by considering all possible classical 1/4 BPS dyons.

Here, we need to make one final consistency check. Not only the value
of the potential ${\cal U}(y)$ at the `minimum' of the effective potential
$U_{\rm eff}(y)$  should yield the right value, which led to the above
identification, but also it should have the `minimum' at the right
value. In other words, we must recover the central relationship $q_A=
\bar{q}_A(y) = L_{AB}(y) a^B$, which is  of field theory origin, by minimizing
the low energy dynamics. The final Hamiltonian with the potential is
\beq
{\cal H} =
 \frac{1}{2} h_{ij}(y)\dot{y}^i \dot{y}^j
  +  \frac{1}{2} (L^{-1})^{AB}(y) q_A q_B
+ \frac{1}{2} a^A L_{AB}(y) a^B,
\eeq
where $\dot{y}^i = h^{ij} (p_j - q_A w^A_{\; j})$.
As in the previous section, we can reexpress this as
\beq
{\cal H} = \frac{1}{2} h_{ij}(y) \dot{y}^i \dot{y}^j
+ \frac{1}{2}(L^{-1})^{AB}(y) (q_A \mp \bar{q}_A(y))(q_B\mp \bar{q}_B(y))
\pm {\cal Z} ,
\eeq
where the central charge is
\beq
{\cal Z} = \bar{q}_A(y)\,(L^{-1})^{AB}(y)\, q_B = a^A \,q_A.
\eeq
The BPS bound is saturated when $\dot{y}^i=0$ and $q_A = \bar{q}_A(y)$,
exactly as we have hoped for. This completes the derivation of the
potential $\cal U$.

The effective Lagrangian for the low energy is the sum of the usual kinetic
term on the moduli space and the potential $\cal U$,
\beq
{\cal L} = \frac{1}{2} \; h_{ij} \dot{y}^i \dot{y}^j + \frac{1}{2}
 L_{AB}  (\dot{\psi}^A + w^A_{\; i} \dot{y}^i) (\dot{\psi}^B+ w^B_{\;
j}\dot{y}^j) -  \frac{1}{2} a^A L_{AB}(y) a^B,
\eeq
or, more compactly,
\beq
{\cal L} = \frac{1}{2}\, g_{\mu\nu} \dot{z}^\mu \dot{z}^\nu -
\frac{1}{2}\, g_{\mu\nu} (a\cdot K)^\mu (a\cdot K)^\nu.
\eeq
Note that, given the Higgs expectation values, the potential is completely
determined in terms of geometry of the monopole moduli space. As showed
by Alvarez-Gaume and Freedman~\cite{alvarez}, and as pointed out by
D. Tong~\cite{tong} very recently,
this form of potential is exactly what one needs to extend the dynamics
to a supersymmetric one. In particular, the triholomorphicity of $a\cdot K$
that follows from gauge invariance, guarantees that the low energy dynamics
will have 8 real supercharges. In this sense the dynamics itself is 1/2 BPS
with respect to the Yang-Mills field theory. The quantum counterpart
of the classical 1/4 BPS dyons should break additional half of these remaining
8 supercharges, and is realized as finite energy BPS states of this low
energy theory itself. In the next section we will explore this supersymmetric
dynamics in some detail.

\section{Supersymmetry and BPS Bound}

We begin with the N=4 supersymmetric quantum extension of the above
effective action~\cite{alvarez}. Its form is similar to the usual
sigma model action but modified by a coupling to the triholomorphic
Killing vector $G\equiv a\cdot K$. The supersymmetric Lagrangian
written with real fermions
is
\beqn
{\cal L}&=&{1\over 2} \biggl( g_{\mu\nu} \dot{z}^\mu \dot{ z}^\nu +
ig_{\mu\nu} \bar\psi^\mu \gamma^0 D_t \psi^\nu + {1\over 6}
R_{\mu\nu\rho\sigma}\bar\psi^\mu \psi^\rho \bar\psi^\nu \psi^\sigma
\biggr.
\nonumber\\
&& \biggl. - g^{\mu\nu} G_\mu G_\nu - D_\mu G_\nu  \bar\psi^\mu
\gamma_5\psi^\nu  \biggr),
\label{action}
\eeqn
where  $\psi^\mu$  is a two-component anticommuting Majorana spinor
and $\gamma^0= \sigma_2$, $\gamma_5= \sigma_3$, and $\bar\psi=\psi^T
\gamma^0$.
The metric here is hyperk\"ahler, endowed with three complex
structures $f^{(a)\mu}\,_\nu (a=1,2,3)$ that satisfy
\beqn
&&f^{(a)}f^{(b)}  = - \delta^{ab} +\epsilon^{abc} f^{(c)}, \\
&& D_\mu f^{(a)\nu}\,_\rho  =0\, ,
\label{complex}
\eeqn
and the Killing vector $G^\mu$ is triholomorphic, i.e., its action
preserves the three complex structures. From now on, we
will use $f$ to denote any one of the three complex structures,
unless noted otherwise.

With vielbein $e_\mu^A$ and the spinors $\psi^A = e_\mu^A \psi^\mu$,
we define supercovariant momenta by
\begin{eqnarray}
&& \pi_\mu \equiv p_\mu -{i\over 2}\omega_{AB\mu}
\bar\psi^A \gamma^0 \psi^B
\label{cov}
\end{eqnarray}
where the  $p$'s are canonical momenta of coordinate $z$'s,
and $\omega^A\,_{B,\mu}$ is the spin connection.  The canonical
commutation relations are $[z^\mu, p_\nu ] = i\delta^\mu_\nu $ and $\{
\psi^A_\alpha, \psi^B_\beta\} = \delta^{AB}\delta_{\alpha\beta}$.
SUSY generators in real form are:
\begin{eqnarray}
&&Q_\alpha = \psi^\mu_\alpha \pi_\mu
+i(\gamma^0\gamma_5 \psi^\mu)_\alpha G_\mu ,\\
&&Q^f_\alpha = f^\mu\,_\nu \psi^\nu_\alpha \pi_\mu
+i(\gamma^0\gamma_5 f^\mu\,_\nu\psi^\nu)_\alpha G_\mu,
\label{generator}
\end{eqnarray}
which satisfy the following SUSY algebra:
\beqn
&&\{Q_\alpha,Q_\beta  \}  =\{Q^f_\alpha,Q^f_\beta\}=2
 \delta_{\alpha\beta} \; {\cal H} +
2 i(\gamma^0\gamma_5)_{\alpha\beta} \; {\cal Z} \\
&& \{Q_\alpha,Q^f_\beta  \} =0
\label{algebra}
\eeqn
Similarly, supercharges associated with different complex
structures $f^{(a)}$ anticommute.
The Hamiltonian $\cal H$ and the central charge $\cal Z$ read
\beqn
&&{\cal H}=
{1\over 2} \biggl( {1\over \sqrt{g}}\pi_\mu \sqrt{g }g^{\mu\nu}\pi_\nu
+ G_\mu G^\mu -{1\over 4}R_{\mu\nu\rho\sigma}\bar\psi^\mu
\gamma^0 \psi^\nu \bar\psi^\rho
\gamma^0 \psi^\sigma + D_\mu G_\nu  \bar\psi^\mu \gamma_5\psi^\nu \biggr)\\
&& {\cal Z}= G^\mu \pi_\mu -{i\over 2}  (D_\mu G_\nu) \bar\psi^\mu
\gamma^0\psi^\nu
\label{hamiltonian}
\eeqn
It is  easily checked that the central charge $\cal Z$ indeed commutes with
all SUSY generators.

For spectrum analysis, SUSY generators in complex form are more useful.
Introducing $\varphi\equiv {1\over \sqrt{2}} (\psi_1^\mu -i\psi_2^\mu)$,
and defining $ Q\equiv {1\over \sqrt{2}} (Q_1-iQ_2)$,
one finds
\beqn
&&Q = \varphi^\mu \pi_\mu
+ i \varphi^{*\mu} G_\mu,\\
&& Q^{\dagger}=\varphi^{*\mu} \pi_\mu
- i \varphi^{\mu} G_\mu,
\label{comgenerator1}
\eeqn
which generates the following simple algebra:
\begin{eqnarray}
&&\{Q,Q^{\dagger}\}=\{Q^f,{Q^f}^{\dagger}\}  =2 {\cal H},\\
&&\{Q,Q\}  = \{Q^f,Q^f\}=-\{Q^{\dagger},Q^{\dagger}\} =
-\{{Q^f}^{\dagger},{Q^f}^{\dagger}\}=
2i{\cal Z},\\
&& \{Q,Q^f\}=\{Q^{\dagger}, Q^f\}=0.
\label{algebra2}
\end{eqnarray}
Again, supercharges associated with different complex
structures $f^{(a)}$ anticommute.

It is easy to read off  the  BPS condition for quantum states that preserves
half of supersymmetries. Depending on the sign of central charge, we find
\begin{eqnarray}
&&(Q\mp iQ^\dagger)|\Phi\rangle  =0,
\label{comgenerator2}
\end{eqnarray}
so that the given state may  saturate the condition ${\cal H}=\pm {\cal Z}$.
We can express this BPS condition in more geometrical fashion by transcribing
the wavefunction to differential forms on the moduli
space~\cite{witten1}. Note that 
\beqn
&& [i\pi_\mu,\varphi^\nu]= -\Gamma^\nu_{\mu\rho} \varphi^{\rho},\\
&& [i\pi_\mu, \varphi^*_\nu]=
\Gamma^\rho_{\mu\nu} \varphi^*_{\rho},\\
&& \{ \varphi^\mu,\varphi^{*}_\nu\}=\delta_\nu^\mu.
\label{comgenerator3}
\end{eqnarray}
Furthermore, the wavefunction has the following general form,
\beqn
&& |\Phi \rangle =
\sum_p \frac{1}{p!}\,\Omega_{\mu_1 \cdots \mu_p}(z^\mu)
\varphi^{\mu_1}\cdots \varphi^{\mu_p}|0\rangle \\
&& \varphi^{*\mu}|0\rangle =0.
\label{comgenerator}
\eeqn
The coefficients $\Omega_{\mu_1 \cdots \mu_p}$ are completely antisymmetric
and may be regarded as those of a $p$-form. In this language where we
interpret $\varphi^\mu$ and $\varphi^*_\mu$ as
a natural cobasis $dz^\mu$ and a natural basis $\partial\over \partial
z^\mu$,  one finds that the following replacement can be made:
\beqn
&& i\varphi^\mu \pi_\mu \rightarrow d \,,
\ \ i\varphi^{*\mu} \pi_\mu \rightarrow -\delta,\\
&& \varphi^{*\mu} G_\mu \rightarrow i_G\,, \ \
 i{\cal Z} \rightarrow {\cal L}_G\equiv di_G +i_G d,
\label{comgenerator4}
\eeqn
where $i_G$ denotes the natural contraction of the vector field $G$
with a differential form, and $\delta$ is the Hermitian conjugate 
of $d$. The BPS equation now becomes
\begin{eqnarray}
(d  - i_G)\, \Omega = \mp\, i(\delta -G\wedge)\Omega
\label{comgenerator5}
\end{eqnarray}
Solving this first order system,
we should recover all 1/2 BPS and 1/4 BPS states of the underlying
Yang-Mills field theory. Work is currently in progress to solve this
equation in the simplest case of $SU(3)$ \cite{piljin}.

\section{Conclusion}

We have found the low energy effective Lagrangian of 1/2 BPS monopoles
in vacua of misaligned Higgs expectation values. This low energy
effective theory produces 1/4 BPS dyons as BPS configurations of the
nonrelativitic Hamiltonian. The kinetic term is given by the usual
moduli space metric of 1/2 BPS monopoles, while the potential term is
also determined by the same geometrical data. Its precise form is
given by one half of the norm of certain Killing vector field, which
allows a supersymmetric extension.

There are several directions to go from here. Our derivation relies
heavily on the established properties of 1/4 BPS dyons and monopole
moduli space. While there is little doubt that this is a valid
derivation, it may be worthwhile to rederive our exact result from a
different perspective. For instance, one may imagine deriving the
exact bosonic potential from a point particle point of view. Another
venue would be to find the supersymmetric low effective  energy Lagrangian
directly from the field theory, along the line of Gauntlett and
Blum~\cite{blum}. Naturally, we expect to recover
the 1/4 BPS dyon spectra as quantum BPS states of this low energy
dynamics. The actual form of the wavefunction is currently under
investigation for the minimal case of $SU(3)$ \cite{piljin}.

Another interesting question concerns monopoles when the symmetry
breaking is not maximal~\cite{ejw3}. The gauge symmetry breaking is
determined by both $\hat{b}\cdot\phi(\infty)$ and $\hat{a}\cdot
\phi(\infty)$. If there is unbroken nonabelian gauge symmetry by
$\hat{b}\cdot\phi(\infty)$, some of magnetic monopoles become
massless. The moduli space acquires an enhanced isometry,
corresponding to unbroken gauge groups. By $\hat{a}\cdot\phi(\infty)$,
the unbroken gauge symmetry could remain unbroken or gets
broken~\cite{kml}. In the former case, the massless monopole clouds
screen the color magnetic charge of massive monopoles, and the
strength of static monopole-monopole force, say in the singlet
channel, may be different from  naive expectation. Such a deviation
has been observed in the large $N$ context quite
recently~\cite{rey}. It would be quite interesting to quantize
massless monopole motion and find the resulting quantum effective
potential between massive monopoles. However, we should also point out
that it is still unclear whether the moduli space dynamics is a valid
approximation in the case of nonmaximal symmetry breaking.

\vspace{3mm}

\centerline{\bf Acknowledgments}

We thank Sangmin Lee for drawing our attention to works related to
Ref.~\cite{alvarez}.  D.B. is supported in part by Ministry of
Education Grant 98-015-D00061.  C.L. and K.L.  are supported in part
by the SRC program of the SNU-CTP and the Basic Science and Research
Program under BRSI-98-2418.  D.B. and K.L. are also supported in part
by KOSEF 1998 Interdisciplinary Research Grant 98-07-02-07-01-5, and
C.L.  by KOSEF Grant 97-07-02-02-01-3.

\end{document}